# Quasiparticle Interference and Strong Electron-Mode Coupling in the Quasi-One-Dimensional Bands of Sr$_2$RuO$_4$

Zhenyu Wang[*1], Daniel Walkup[*2,3], Philip Derry[4], Thomas Scaffidi[5], Melinda Rak[1], Sean Vig[1], Anshul Kogar[1], Ilija Zeljkovic[2], Ali Husain[1], Luiz H. Santos[6], Yuxuan Wang[6], Andrea Damascelli[7], Yoshiteru Maeno[8], Peter Abbamonte[1], Eduardo Fradkin[6], and Vidya Madhavan[1]

[1]*Department of Physics and Frederick Seitz Materials Research Laboratory, University of Illinois Urbana-Champaign, Urbana, Illinois 61801, USA*

[2]*Department of Physics, Boston College, Chestnut Hill, Massachusetts 02467, USA*

[3]*National Institute of Standards and Technology, Gaithersburg, Maryland 20899, USA*

[4]*Department of Chemistry, Physical & Theoretical Chemistry, Oxford University, South Parks Road, Oxford OX1 3QZ, United Kingdom.*

[5] *Rudolf Peierls Centre for Theoretical Physics, Oxford OX1 3NP, United Kingdom*

[6] *Department of Physics and Institute for condensed Matter Theory, University of Illinois at Urbana-Champaign, Urbana, Illinois 61801, USA*

[7]*Department of Physics & Astronomy, University of British Columbia, Vancouver British Columbia, V6T 1Z1, Canada, Quantum Matter Institute, University of British Columbia, Vancouver, British Columbia V6T 1Z4, Canada*

[8]*Department of Physics, Graduate School of Science, Kyoto University, Kyoto 606-8502, Japan*

*These authors contributed equally to this work

**The single-layered ruthenate $Sr_2RuO_4$ has attracted a great deal of interest as a spin-triplet superconductor with an order parameter that may potentially break time reversal invariance and host half-quantized vortices with Majorana zero modes. While the actual nature of the superconducting state is still a matter of controversy, it has long been believed that it condenses from a metallic state that is well described by a conventional Fermi liquid. In this work we use a combination of Fourier transform scanning tunneling spectroscopy (FT-STS) and momentum resolved electron energy loss spectroscopy (M-EELS) to probe interaction effects in the normal state of $Sr_2RuO_4$. Our high-resolution FT-STS data show signatures of the β-band with a distinctly quasi-one-dimensional (1D) character. The band dispersion reveals surprisingly strong interaction effects that dramatically renormalize the Fermi velocity, suggesting that the normal state of $Sr_2RuO_4$ is that of a `correlated metal' where correlations are strengthened by the quasi 1D nature of the bands. In addition, kinks at energies of approximately 10meV, 38meV and 70meV are observed. By comparing STM and M-EELS data we show that the two higher energy features arise from coupling with collective modes. The strong correlation effects and the kinks in the quasi 1D bands may provide important information for understanding the superconducting state. This work opens up a unique approach to revealing the superconducting order parameter in this compound.**

The electronic properties of complex oxides are highly sensitive to electron-electron interactions as well as interactions of electrons with other collective modes[1-4]. Identifying these many-body effects is critical to understanding the driving forces behind many of their exotic phases. The unconventional p-wave superconductor $Sr_2RuO_4$ in particular, is a fundamentally interesting material system[5-7]. Understanding the nature of its unique superconducting state with spin-triplet pairing symmetry requires an intimate knowledge of its normal state properties[8]. Yet, the effects of interactions in the normal state of this system are yet to be sorted out, with the conventional belief that it is a Fermi liquid conflicting with reports of large band dependent mass renormalizations and strong correlation effects[8-12]. Much of this uncertainty can be traced to the multiband nature, with the competition between 1D and 2D, the interplay between spin and lattice degrees of freedom[13], as well as the strong k-dependent spin-orbital entanglement of the normal state wavefunction, which makes the description of superconductivity in terms of pure spin-triplet (and/or singlet) eigenstates questionable[14].

The overall band structure of $Sr_2RuO_4$ is well known[8, 15] and consists of three Fermi surface (FS) sheets with distinct characteristics (Fig. 1a). The three bands are primarily derived from the ruthenium $4dt_{2g}$ orbitals. Hybridization between $d_{xz}$ and $d_{yz}$ orbitals leads to two sets of quasi-1D FS sheets. One is the hole-like α sheet near X and the other is the electron-like β sheet near Γ. The in-plane $d_{xy}$ orbital on the other hand forms the electron-like, quasi-2D γ sheet centered at Γ. The dominant superconducting instability can be placed either on the γ band or the α and β bands in different theoretical approaches[16-20]. Consequently, the symmetry of the resulting superconducting state and the pairing `glue' are still unclear. Thus, it is essential to distinguish the interaction effects on the different bands, which may generate or assist the superconducting pairing in this material. Angle-resolved photoemission spectroscopy (ARPES) data have revealed kinks at energies of 40meV, 50-60meV and 70-80meV in the dispersion of quasi-2D γ band[4, 13, 21]. While there have been some discussions of self energy effects for the α band[21], no sharp features have been unambiguously identified on the quasi-1D bands [13, 21, 23].

In this work we study the effects of interactions on the electronic structure of $Sr_2RuO_4$ using the complementary techniques of FT-STS and meV resolution M-EELS. FT-STS is a powerful tool to study electron behavior both in *r*-space and *k*-space simultaneously, and has been successfully used to study the nanoscale spectroscopic properties of high $T_C$ superconductors[24], heavy fermion systems[25, 26] and the bilayer strontium ruthenate[27]. M-EELS on the other hand is a powerful technique for measuring the energy and dispersion of collective excitations that couple strongly to electrons[28]. So far, neither of these techniques has been successfully applied to $Sr_2RuO_4$. Here, we use FT-STS to visualize the quasiparticle interference in the normal state and determine band dispersion with high precision which allows us to determine correlation effects as well as the effects of lattice and spin excitations on the electronic structure. Our FT-STS data reveal that the β band displays signatures of quasi 1D behavior with a dispersion that reveals a dramatic suppression of Fermi velocity. We further find a low energy suppression of the density of states centered approximately at the Fermi energy. These combined observations suggest that quasi 1D band character accentuates correlation effects, which in turn has important consequences for the electronic properties of the quasi 1D bands in this system. We note that while there are many studies of correlation effects in pure 1D systems, the effects on quasi 1D bands are less well known. At higher energies we find kinks in the dispersion which are also observed in our M-EELS data suggesting that they originate from the coupling of quasiparticles with collective bosonic modes such as phonons. The strong correlation effects and the identification of energy scale of the kinks in the quasi 1D bands may provide key information needed to obtain a microscopic model for the superconducting state. Moreover, our success in obtaining high quality data using FT-STS for the first time provides a new pathway for exploring the quasiparticles below $T_c$ which would reveal the momentum dependence of the superconducting energy gap, $\Delta(\boldsymbol{k})$ and help distinguish the pairing mechanism in $Sr_2RuO_4$.

$Sr_2RuO_4$ has a layered perovskite structure similar to cuprate superconductors[5] (Fig. 1b). Cleaving could in principle expose two kinds of natural non-polar surfaces, either SrO or $RuO_2$ planes[18], although cleaving at the SrO plane is thought to be more likely. The topographic features can be highly dependent on the cleave temperature[15, 29]. $Sr_2RuO_4$ single crystals studied here were cleaved at ~ 80K *in situ* and then transferred to a scanning tunneling microscope (STM) held at 4.3K. Figure 1d shows a typical topographic image obtained on the cleaved surface, showing a square lattice with atomic spacing ~ 3.9 Å. The 6.3 Å atomic step height seen near this scan range (supplementary information part *I*) suggests a preferential termination layer, which we believe to be the SrO plane. The bright protrusions in the STM image are most likely Sr atoms[30] while the impurities that look like dark crosses in our topography can be tentatively assigned to CO adsorbates[31]. Although a secondary modulation is almost invisible in the topographic images, their Fourier transforms show additional peaks at $\sqrt{2} \times \sqrt{2}$ positions arising from the ($\sqrt{2} \times \sqrt{2}$ )R45° surface reconstruction seen in low energy electron diffraction[30] and ARPES measurments[15]. Our energy-integrated M-EELS data taken along the (H, H) direction in reciprocal space also shows a peak near (1/2, 1/2), corresponding to this reconstruction (supplemental Fig. S2). A schematic of this reconstruction is shown in Fig. 1c. As a consequence of this reconstruction, the first Brillouin Zone (BZ) is reduced to half (dashed black square in Fig. 1a), which gives rise to band folding with respect to the ($\pi/a_0$, 0)- (0, $\pi/a_0$) line. This effect plays an important role in the quasiparticle interference (QPI) pattern, which we will return to in detail.

A typical differential conductance spectrum i.e., *dI/dV* (*r*, eV) is shown in Fig. 1e. This spectrum is representative of the sample: since the native impurity concentration is low, the density of states is quite homogeneous. There are features at approximately 38meV on both sides of the Fermi energy ($E_F$), which is similar to earlier data on Ti-doped $Sr_2RuO_4$ samples[32]. The particle-hole symmetric nature of the 38meV peak can be clearly seen in the derivative of the spectrum (Fig. S3) and suggests that it originates from coupling with a collective mode[33]. In fact we find collective modes at this energy in both the STM derived dispersion and M-EELS data and we will return to this energy scale later. Interestingly, similar to the spectra measured below $T_c$ in previous work[32], our data show a low energy gap-like feature with an energy scale of approximately 10 meV. While the origin of this anomaly is unclear, one possibility is that it represents a suppression of the tunneling density of states of the α and β bands due to accentuated correlation effects arising from their quasi-1D nature[34] as discussed later in this paper. Further temperature dependent studies would be necessary to elucidate the origin of this feature.

We now apply the technique of FT-STS to $Sr_2RuO_4$. In FT-STS the spatial modulation arising from the elastic scattering of quasiparticles can be measured as *dI/dV (r, eV)* maps and Fourier-transformed to extract scattering vectors (Q-vectors), which connect k-space electronic states under some selection rules. High-resolution FT-STS measurements not only allow us to measure details of the band dispersion, but can also be used to reveal orbital/spin textures. Figures 2a-c show representative *dI/dV (r, eV)* maps on $Sr_2RuO_4$ at a few different energies. The Fourier transforms of of *dI/dV* maps demonstrating a rich array of scattering channels in this material are shown in Fig. 2d-l, where a sequence of inequivalent sets of scattering channels are labeled as $q_i$: i=0,1,2,3,4. We find that $q_0$ is rather non-dispersive. While this feature will be discussed in further detail elsewhere we note that it potentially originates from Friedel oscillations generated by impurities whose signatures are unusually strong due to the quasi-1D nature of the bands. Here we focus on the dispersing channels $q_1$-$q_4$. We note that although for ease of discussion we focus on the scattering vectors along the high symmetry directions (π, 0) and (π, π), there is of course a continuous array of scattering channels, which reflect the Fermi surface topology. These are marked as arcs in the appropriate colors in Fig. 2 with the understanding that all the scattering vectors along arcs of a particular color originate from a set of scattering processes connecting two particular bands.

Identification of the origin of the Q-vectors requires comparison between the band structure (Fig. 3a), predicted QPI (Fig. 3c and d) and the measured data (Fig. 2 and Fig. 3b). The detailed analysis of the Q-vectors is presented in the supplemental information part *II*, and here we only present a summary. The dominant signal in the FT is along (π, 0) direction, labeled $q_1$. A complete analysis of this feature indicates that $q_1$ represents the intra-β-band scattering (pink arrows, also shown as pink arcs in Fig. 2). The corresponding umklapp scattering process, labeled as $q_1'$, is also visible in our data (dashed pink arrows). These two scattering processes are illustrated in figure 3a.

To understand the remaining Q-vectors we require three essential pieces of information. First, ARPES data show two β-bands, one identified as the bulk β-band and the other as a surface β-band attributed to rotated $RuO_6$ octahedra at the surface[35, 36]. In our data we also observe a second dispersing square feature (orange arrows, also shown as orange arcs in Fig. 2) that emerges close to Fermi level (Fig. 2g). As discussed in detail after obtaining the dispersion, both the qualitative and quantitative behavior of $q_1$ and $q_2$ suggest a common origin with the two β-ARPES bands. We note that $q_2$ may in principle originate from the α band ($q_α$; gray arrows in Fig. 3c and d) but the concave curvature of $q_2$ is identical to the curvature of $q_1$ (see Fig. 2i for example) and quite different from the convex curvature expected of $q_α$ (Fig. 3c and d) which indicates that this is unlikely.

Second, most of the γ band (except around ($π/a_0$, $π/a_0$)) is composed of the planar $d_{xy}$ orbital. The matrix elements for the coupling of an STM tip to planar orbitals are typically small. Correspondingly, the signal from this band is therefore mostly absent from our data[27]. Around ($π/a_0$, $π/a_0$) however, a square-shaped feature $q_3$ appears, which can be distinguished from $q_2$ by a discontinuity as we trace the contours of β band. This signal potentially originates from small portion of the $γ$ band along ΓX direction that acquires $d_{xz}/d_{yz}$ orbital character due to hybridization[14], thereby making it visible to STM measurements.

Third, the rotation of the oxygen octahedra at the surface creates band folding at the surface. This has a distinct effect on the QPI as seen in the comparison of theoretical calculations of the QPI with (Fig. 3d) and without band-folding (Fig. 3c). One of the effects of band folding is the appearance of parallel lines inside the yellow rectangle in figure 3d. These folded features are labeled $q_4$ and can be seen clearly in our data (Fig.3b, Fig 2h-l and SI part *II*).

Having understood the dominant features in the Fourier transforms, we turn to their dispersions. Linecuts of the FTs were obtained in the two high symmetry directions, (π, 0) (corresponding to scattering vectors in the ΓM direction) and (π, π) (ΓX direction). The position of each peak in the linecut was determined by with Gaussian fitting function on a linear background (SI part *III*). The peak positions representing the energy and momentum resolved Q-vectors are plotted in Fig. 4a. This plot does not include $q_4$ since it represents a folded band and is not an independent Q-vector. By comparing the shapes of $q_1$ and $q_2$ in the FTs shown in Fig. 2 as well as their dispersion, it is apparent that $q_2$ and $q_1$ are related to each other with one branch being shifted in momentum with respect to the other. In essence $q_2$ and $q_1$ behave like two versions of the β band. While there are a few possible explanations for such a splitting such as surface induced spin-orbit effect or magnetic fluctuations, as mentioned earlier, a secondary β band was also observed in ARPES measurements[35, 36] where the two bands were attributed to a surface β band arising from the rotated $RuO_6$ octahedra at the surface and the bulk β band, respectively. Quantitatively, the two Fermi wavevectors ($k_F$) observed by us are 0.62+-0.02 and 0.68+-0.02 (in units of $\pi/a_0$). These values are identical those observed in multiple ARPES experiments[35-37] indicating that the STM and ARPES bands have the same origin. ARPES studies identified the band with the larger $k_F$ ($q_1$) to be the bulk branch while the band with the smaller $k_F$ ($q_2$) was identified as a surface branch. Interestingly, we find support for this scenario by looking at band folding effects. We find that the band identified by ARPES to be a surface band ($q_2$) shows clear folding (Fig. 2h-l) as evidenced by the presence of $q_4$, while the band folding is non-existent for $q_1$ below the Fermi energy as seen in Fig. 2d-g. The comparison with ARPES, as well as the differences in band folding effects for $q_1$ and $q_2$ suggest that we measure both the surface and bulk bands by STM with $q_1$ being identified as the bulk band.

A striking feature of the dispersion of the β band is the change in slope of $q_1$ and $q_2$ near $E_F$, indicating a strong renormalization of the Fermi velocity in this band (Fig. 4a). From our data, the Fermi velocity of $q_1$ is renormalized to 0.46 eV Å. This is smaller by a factor of 1.4 compared to the values obtained from de Haas-van Alphen oscillations. We believe that the discrepancy with dHvA indicates a momentum dependent renormalization. A more detailed comparison between STM, dHvA, and ARPES data is present in SI PART *III*. In essence, our high-resolution data reveals that the correlation effects on the β band are much larger than previously thought.

In fact, the quasi 1D nature of the β band may play an important role in strengthening correlation effects. Strong quasi-1D signatures of the β band are seen in many facets of our data. First, tracking the β band contours shown in Fig. 2, we see that the band is remarkably flat for much of its extension. Second, the brightness of the $q_1$/ $q_2$ scattering vectors in the ΓM direction (supplemental Fig. S7) indicate that there is a singularity in the numbers of scattering processes with the same magnitude in this direction. The quasi 1D nature is also reflected in the momentum dependent nature of the renormalization suggested by comparison with dHvA in the previous paragraph. In general, quasi 1D electronic states are expected to share many key features of true 1D systems including non-Fermi liquid behaviors such as a suppression of the one-particle density of states, a large downward renormalization of the Fermi velocity and a large enhancement of the charge and spin susceptibilities at $2k_F$ (Ref. 34). In quasi 1D systems however, these features are rounded by an eventual crossover to 2D Fermi liquid behavior[38, 39]. Many of the predicted effects of 1D bands on the electronic structure are borne out by the STM data presented here, including a zero-bias anomaly in the tunneling DOS at an energy scale of about 10meV and a downward renormalization of the Fermi velocity over the same energy scale. These observations taken together support the scenario of enhanced correlation effects on the β band bolstered its quasi 1D nature. Our observations also suggest that this system may be close to a charge and/or spin density wave instability, as implied by earlier neutron scattering data[40].

In addition to interaction effects near $E_F$, the dispersion in Fig. 4a reveals clear kinks at multiple energy scales. Kinks are ubiquitous in many correlated electron systems and reflect self-energy renormalizations, which carry important information about the effective interactions. Among high-$T_C$ superconductors for example, kinks in the dispersion and have been observed in cuprates[2, 3] and Fe-based superconductors[41]. However, simply observing kinks in the dispersion is typically not sufficient to understand their origin since both electron-electron as well as electron boson interactions may result in kinks[1]. In the following discussion, we employ the complementary techniques of STM and M-EELS to not only identify the energy scales of the kinks but also their potential origin. Kinks were observed in the STM data at energies of ~35meV ($\omega_1$) and ~70meV ($\omega_2$) in the dispersion of $q_1$, and one at about ~32meV above Fermi level for $q_2$. These energy scales are also clearly visible in the extracted self-energy Re$\Sigma(\mathbf{k}, \omega)$ shown in Fig. 4b. The effective self-energy Re$\Sigma(\mathbf{k}, \omega)$ was obtained by subtracting a bare band from the observed dispersion. To avoid artifacts, the bare band was simply chosen to be a straight line connecting two points of q($\omega$=0) and q($\omega$=±110 meV). The details of the bare band used would be very important if one were trying to obtain quantitative information on the self-energy[22]. However, this treatment is sufficient for our analysis where we concentrate on the peak positions. The peak energies of $\omega_1$ and $\omega_2$ in extracted Re$\Sigma(\mathbf{k}, \omega)$ are ~37meV and ~73meV. To further pin down the energy scales and the potential origin of these kinks, we performed M-EELS measurements on the same samples. The results are shown in Fig. 4c. Interestingly M-EELS data predominantly shows the same two peaks, one near 38meV (corresponding to $\omega_1$ in the STM data) and the other at 71meV (corresponding to $\omega_2$). $\omega_1$ disperses with momentum in a manner consistent with an optical phonon. $\omega_2$ however shows anomalous momentum dependence disappearing abruptly as we move away from high symmetry points suggestive of a surface phonon merging with a bulk band. In essence, the comparison of the STM data with M-EELS provides unambiguous evidence that $\omega_1$ and $\omega_2$ arise from collective bosonic modes that strongly couple with the quasi-1D β band in $Sr_2RuO_4$. We note that similar modes were observed in ARPES studies of the 2D γ band[4, 13]. The similarity of these energy scales to the M-EELS data provides strong evidence that the ARPES kinks at these energies also arise coupling of quasiparticles with the same phonons. Interestingly, the dispersion of $q_3$ along the ΓX direction (inset in Fig. 4a) shows an additional kink near 10meV ($\omega_3$). The background subtracted $q_3$ shown in Fig. 4b also clearly reveals $\omega_3$. If one regards $q_3$ as the intra-$\gamma$-band scattering, $\omega_3$ may be related to coupling with the $\Sigma_3$ phonon which exhibits a sharp drop near the zone boundary (~1.9THz) and potentially enhances ferromagnetic (FM) spin fluctuations[30].

The order parameter (OP) of superconductivity and the associated gap structure in momentum space have been long-standing issues in $Sr_2RuO_4$. While compelling experimental evidence favors an odd parity (triplet) state and there is strong evidence that the SC order spontaneously breaks time reversal symmetry[5, 6] suggesting a chiral p-wave state, this latter question has remained controversial primarily due to the possible different roles played by the quasi-1D and 2D bands. The most direct way to distinguish between the various predictions involving pairing on either the quasi-1D bands[17] ($\alpha$, $\beta$) or the quasi-2D band[16] ($\gamma$) is to determine the momentum dependent OP, a task not yet done due to the low superconducting transition temperature of ~1K, sub-meV magnitude of the gap, and the subtle multiband nature. However, our present work gives a unique approach to directly confirm the gap structure on the 1D bands. In the superconducting state, if one places the dominant gap on 1D bands, the $\beta$ band would be gapped out and the CECs of Bogoliubov quasiparticles would form around the nodes near ($\pi$, $\pi$) (ref. 18). As the result, the BQPI pattern, which is dominated by scattering between the ends of these banana-shaped CECs, will change dramatically compared to the normal state pattern (see Fig.5 in ref. 18). Detailed information about $\Delta(k)$ on the $\beta$ band can then be obtained by tracking the energy evolution of this pattern, which will increase our understanding of the pairing symmetry as well as the microscopic pairing mechanism in this material. Finally, the results presented in this paper suggest that the quasiparticle states of the $\beta$ band have a quasi-1D character and are strongly affected by electron interaction effects, thus raising doubts on the picture that the normal state is simply a weakly-correlated Fermi liquid metal. Additional experiments, including the effects of temperature and magnetic fields, will be needed to further elucidate this picture.

## Methods

High-quality $Sr_2RuO_4$ single crystals for the data shown in main text were grown at Kyoto University. The samples were cleaved at liquid Nitrogen temperature and all the dI/dV measurements were taken at 4.3K using a standard lock-in technique with 5meV peak to peak modulation at a frequency of 987.5Hz. Lawler-Fujita drift- correction algorithm is used for the FT-QPI data to remove the drift effects[42]. The results were reproduced on other samples grown at UBC with different tungsten tips (see Supplementary Information Part *III*).

M-EELS measures the bosonic, density response function, $\chi''(q, \omega)$, of a material surface[28]. These experiments were carried out at a beam energy of 50 eV on crystals of $Sr_2RuO_4$, cleaved under vacuum and subsequently cooled to 100 K. Elastic scattering from the (1, 0) and (1, 1) Bragg reflections were used *in situ* to construct an orientation matrix translating between diffractometer angles and reciprocal space. The Miller indices (H, K) designate the transferred momentum in tetragonal units, such that $\mathbf{q} = 2\pi(H, K)/a$, where a = ~3.9 Å is the in-plane lattice parameter.

## Acknowledgements

We thank Ziqiang Wang, Hsin Lin, Seamus Davis, and Steve Kivelson for useful conversations. V.M. gratefully acknowledges funding from the US Department of Energy, Scanned Probe Division under Award Number DE-SC0014335f or the support of Z. W., D.W., and I.Z. P.A. and V.M acknowledge the Gordon and Betty Moore Foundation Grant No. GBMF4860. The theoretical work was supported in part by the Gordon and Betty Moore Foundation's EPiQS Initiative through Grant No. GBMF4305 at the Institute for Condensed Matter Theory of the University of Illinois (L.S. and Y.W.), and by a grant of the National Science Foundation No. DMR1408713 at the University of Illinois (E.F.). M-EELS experiments were supported by the Gordon and Betty Moore Foundation's EPiQS Initiative through Grant GBMF4542. T.S. acknowledges the financial support of the Clarendon Fund Scholarship, the Merton College Domus and Prize Scholarships, and the University of Oxford.

## Author Contributions

Z.W. and D.W. contributed equally to this work. Z.W., D.W., and V.M. designed the STM experiments, analyzed the data and wrote the paper. STM experiments were performed by D.W., Z.W., and I.Z. Y.M. was responsible for single crystal growth and structural analysis. A.D. helped with conceiving the experiment, data analysis and comparison with ARPES. E.F., L. S. and Y.W. conceived the theoretical explanation for this work. P.D., and T.S. performed analytical model calculations. M.R., S.V., A.K., A.H., and P.A., were involved in the M-EELS studies.

## Competing financial interests

The authors declare no competing financial interest.

## Corresponding author

Correspondence to: Vidya Madhavan(vm1@illinois.edu)

# References

1. Byczuk, K. *et al.* Kinks in the dispersion of strongly correlated electrons. *Nature Phys.* **3**, 168- 171 (2007).
2. Carbotte, J. P., Timusk, T. & Hwang, J. Bosons in high-temperature superconductors: an experimental survey. *Rep. Prog. Phys.* **74**, 066501 (2011).
3. Lanzara, A. *et al.* Evidence for ubiquitous strong electron-phonon coupling in high-temperature superconductors. *Nature* **412**, 510–514 (2001).
4. Aiura, Y. *et al.* Kink in the dispersion of layered Strontium Ruthenates. *Phys. Rev. Lett.* **93**, 117005 (2004).
5. Mackenzie, A. P. & Maeno, Y. The superconductivity of $Sr_2RuO_4$ and the physics of spin-triplet pairing. *Rev. Mod. Phys.* **75**, 657 (2003).
6. Kallin, C. Chiral *p*-wave order in $Sr_2RuO_4$. *Rep. Prog. Phys.* **75**, 042501 (2012).
7. Jang, J. *et al.* Observation of half-height magnetization steps in $Sr_2RuO_4$. *Science* **331**, 186-188 (2011).
8. Bergemann, C. *et al*. Quasi-two-dimensional Fermi liquid properties of unconventional superconductor $Sr_2RuO_4$. *Adv. Phys.* **52**, 639–725 (2003).
9. Liebsch, A. & Lichtenstein, A. Photoemission quasiparticles spectra of $Sr_2RuO_4$. *Phys. Rev. Lett.* **84**, 1591 (2000).
10. Kidd, T. E. *et al.* Orbital dependence of the Fermi liquid state in $Sr_2RuO_4$. *Phys. Rev. Lett.* **94**, 107003 (2005).
11. Stricker, D. *et al.* Optical response of $Sr_2RuO_4$ reveals universal Fermi-Liquid scaling and quasiparticles beyond Laudau theory. *Phys. Rev. Lett.* **113**, 087404 (2014).
12. Zhang, G., Gorelov, E., Sarvestani, E. & Pavarini, E. Fermi surface of $Sr_2RuO_4$: spin-orbit and anisotropic Coulomb interaction effects. *Phys. Rev. Lett.* **116**, 106402 (2016).
13. Iwasawa, H. *et al.* Interplay among Coulomb interaction, spin-orbital interaction, and multiple electron-Boson interactions in $Sr_2RuO_4$. *Phys. Rev. Lett.* **105**, 226406 (2010).
14. Veenstra, C. N. *et al.* Spin-orbital entanglement and the breakdown of singlets and triplets in $Sr_2RuO_4$ revealed by Spin- and Angle-Resolved Photoemission Spectroscopy. *Phys. Rev. Lett.* **112**, 127002 (2014).
15. Damascelli, A. *et al.* Fermi surface, surface states, and surface Reconstruction in $Sr_2RuO_4$. *Phys. Rev. Lett.* **85**, 5194 (2000).
16. Rice, T. M. & Sigrist, M. $Sr_2RuO_4$: A electronic analogue of $^3$He? *J. Phys: Condens. matter* **7**, L643 (1995).
17. Raghu, S., Kapitulnik, A. & Kivelson, S. A. Hidden quasi-one-dimensional superconductivity in $Sr_2RuO_4$. *Phys. Rev. Lett.* **105**, 136401 (2010).
18. Firmo, I. A. *et al.* Evidence from tunneling spectroscopy for a quasi-one-dimensional origin of superconductivity in $Sr_2RuO_4$. *Phys. Rev. B* **88**, 134521 (2013).
19. Wang, Q. H. *et al.* Theory of superconductivity in a three-orbital model of $Sr_2RuO_4$. *Euro. Phys. Lett.* **104**, 17013 (2013).
20. Scaffidi, T., Romers, J. C. & Simon, S. H. Pairing symmetry and dominant band in $Sr_2RuO_4$. *Phys. Rev. B* **89**, 220510(R) (2014).
21. Iwasawa, H. *et al.* Orbital selectivity of the kink in the dispersion of $Sr_2RuO_4$. *Phys. Rev. B* **72**, 104514 (2005).
22. Ingle, N. J. *et al.* Quantitative analysis of $Sr_2RuO_4$ angle-resolved photoemission spectra: Many-body interactions in a model Fermi liquid. *Phys. Rev. B* **72**, 205114 (2005).

23. Kim, C. *et al.* Self-energy analysis of multiple-bosonic mode coupling in $Sr_2RuO_4$. *J. Phys. Chem. Solids.* **72**, 556-558 (2011).
24. Hoffman, J. E. *et al.* Imaging quasiparticle interference in $Bi_2Sr_2CaCu_2O_{8+\delta}$. *Science.* **297**, 1148–1151 (2002).
25. Allan, M. P. *et al.* Imaging Cooper pairing of heavy fermions in $CeCoIn_5$. *Nature Phys.* **9**, 468–473 (2013).
26. Zhou, B. B. *et al.* Visualizing nodal heavy fermion superconductivity in $CeCoIn_5$. *Nature Phys.* **9**, 474–479 (2013).
27. Lee, J. *et al.* Heavy *d*-electron quasiparticle interference and real-space electronic structures of $Sr_3Ru_2O_7$. *Nature Phys.* **5**, 800–804 (2009).
28. Kogar, A., Vig, S., Gan, Y. & Abbamonte P. Temperature-resolution anomalies in the reconstruction of time dynamics from energy-loss experiments, *J. Phys. B: At. Mol. Opt. Phys.* **47**, 124034 (2014).
29. Pennec, Y. *et al.* Cleaving-Temperature Dependence of Layered-Oxide Surfaces. *Phys. Rev. Lett.* **101**, 216103 (2008).
30. Matzdorf, R. *et al.* Ferromagnetism stabilized by lattice distortion at the surface of the *p*-wave superconductor $Sr_2RuO_4$. *Science* **289**, 746 (2000).
31. Stöger, B. *et al.* High chemical activity of a Perovskite surface: Reaction of CO with $Sr_3Ru_2O_7$. *Phys. Rev. Lett.* **113**, 116101 (2014).
32. Barker, B. I. *et al.* STM studies of individual Ti impurity atoms in $Sr_2RuO_4$. *Physica B* **329B**, 1334 (2003).
33. Hlobil, P. *et al.* Tracing the electronic pairing glue in unconventional superconductors via Inelastic Scanning Tunneling Spectroscopy. Preprint at https://arxiv.org/abs/1603.05288 (2016).
34. Fradkin, E., *Field Theories of Condensed Matter Systems* (Cambridge University Press, 2013)
35. Veenstra, C. N. *et al.* Determining the surface-to-bulk progression in the normal-state electronic structure of $Sr_2RuO_4$ by Angle-Resolved Photoemission and density functional theory. *Phys. Rev. Lett.* **110**, 097004 (2013).
36. Liu, S. Y. *et al.* Fermi surface sheet-dependent band splitting in $Sr_2RuO_4$ revealed by high-resolution angle-resolved photoemission spectroscopy. *Phys. Rev. B* **86**, 165112 (2012).
37. Zabolotnyy, V. B. *et al.* Surface and bulk electronic structure of the unconventional superconductor $Sr_2RuO_4$: unusual splitting of the β band. *New J. Phys.* **14**, 063039 (2012).
38. Carlson, E. W., Orgad, D., Kivelson, S. A. & Emery, V. J. Dimensional crossover in quasi-one-dimensional and high-$T_C$ superconductors. *Phys. Rev. B* **62**, 3422 (2000).
39. Biermann, S., Georges, A., Lichtenstein, A. & Giamarchi, T. Deconfinement transition and Luttinger to Fermi liquid crossover in quasi-one-dimensional systems. *Phys. Rev. Lett.* **87**, 276405 (2001).
40. Sidis, Y. *et al.* Evidence for incommensurate spin fluctuations in $Sr_2RuO_4$. *Phys. Rev. Lett.* **83**, 3320 (1999).
41. Allan, M. P. *et al.* Identifying the 'fingerprint' of antiferromagnetic spin fluctuations in iron pnictide superconductors. *Nature Phys.* **11**, 177–182 (2015).
42. Lawler, M. J. *et al.* Intra-unit-cell electronic nematicity of the high-$T_C$ copper-oxide pseudogap states. *Nature* **466**, 347-351 (2010).

# Figure 1

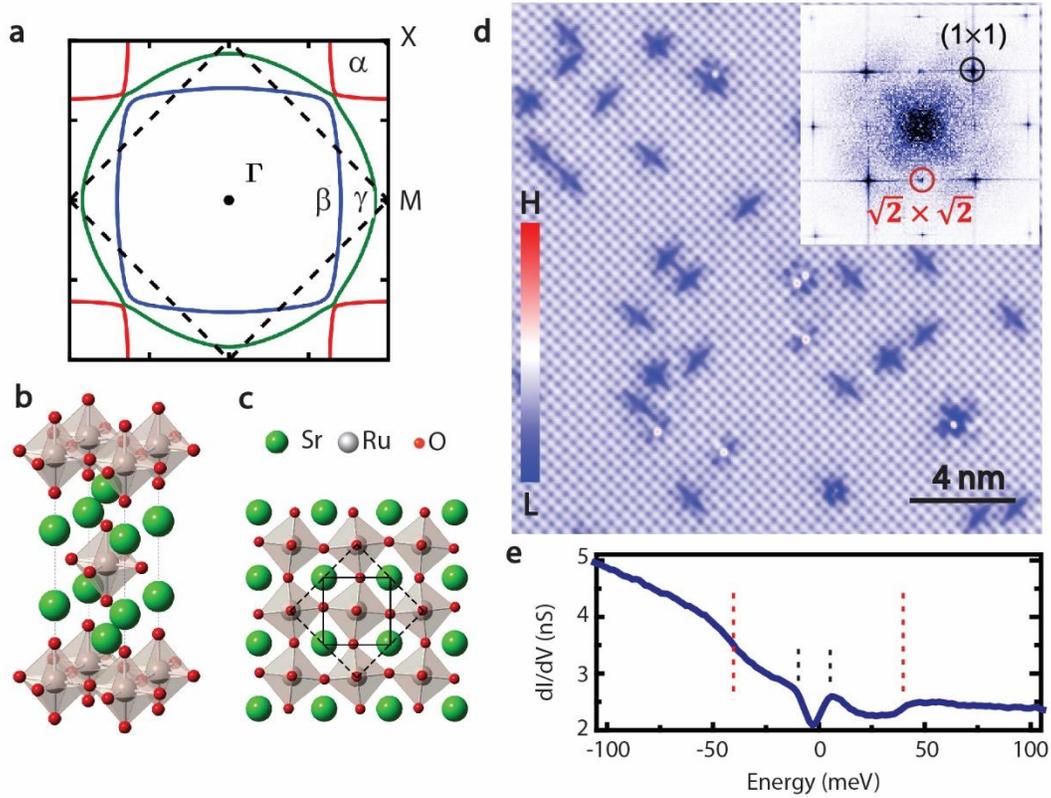

**Figure 1. Fermi surfaces and crystal structure of Sr$_2$RuO$_4$. a,** Bulk Fermi surfaces of Sr$_2$RuO$_4$ calculated with tight binding model. Dashed lines denote the new Brillouin Zone caused by $\sqrt{2} \times \sqrt{2}$ surface reconstruction. **b,** Crystal structure of Sr$_2$RuO$_4$ showing the Ru-centered octahedra. **c,** A schematic top view of the surface reconstruction with rotated RuO$_6$ octahedra. The unit cells with and without rotation are denoted by dashed black square and solid black square, respectively. **d,** Topographic image of Sr$_2$RuO$_4$ showing a usual uniform and square lattice with spacing of ~3.9 Å between atoms (bias voltage V$_S$=70mV, tunneling current I$_t$=100pA). The inset shows its Fourier transform: Black circle represents Bragg peak and red circle for the $\sqrt{2} \times \sqrt{2}$ reconstruction peak. **e,** Typical differential conductance spectrum taken in defect-free region (I$_t$=265pA, V$_S$=110mV). The red and black dashed lines denote two features with energy scales approximately 38meV and 10meV, respectively.

# Figure 2

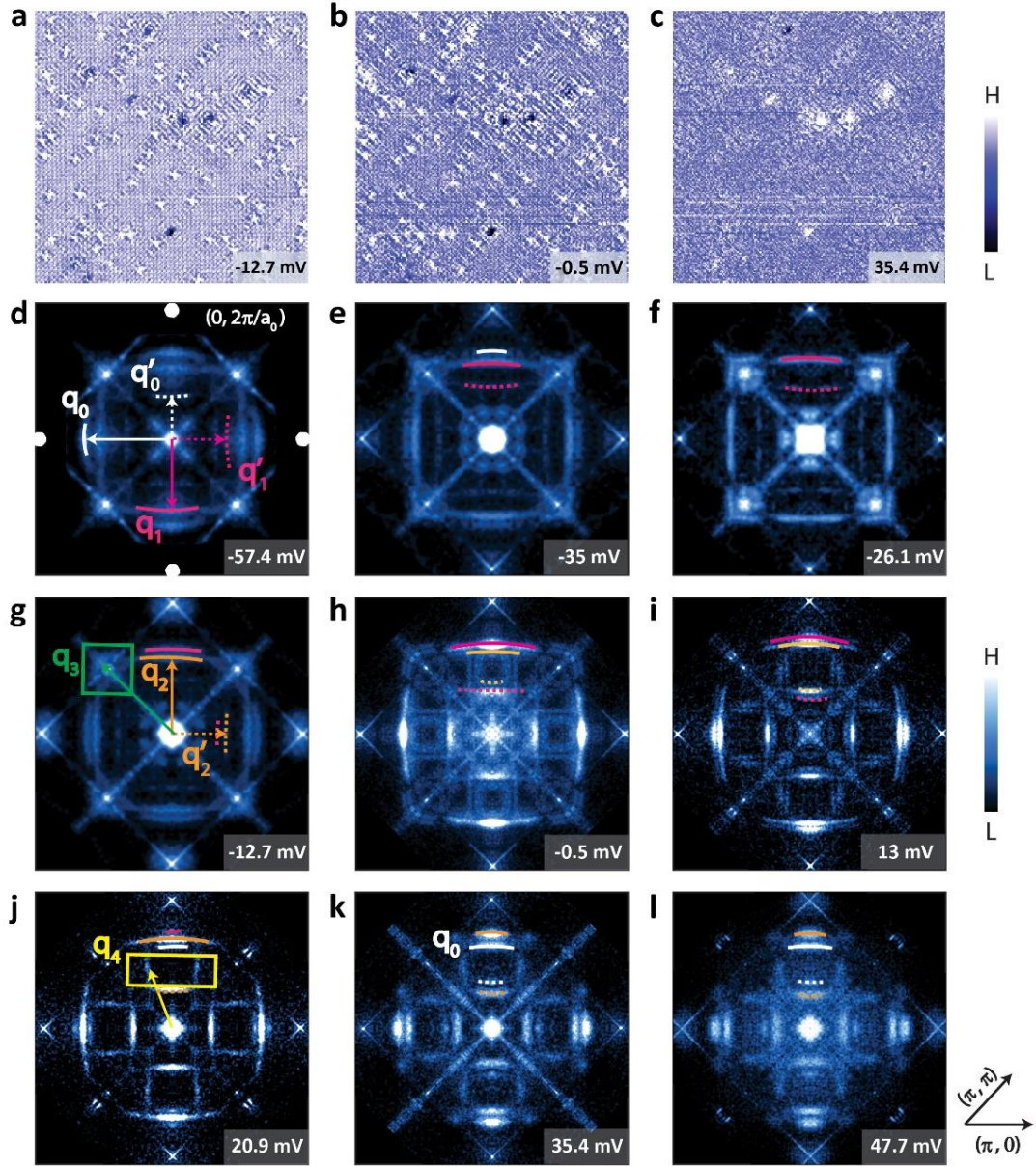

**Figure 2. Quasiparticle interference (QPI) of Sr$_2$RuO$_4$. a-c**, Spatially resolved dI/dV conductance maps at -12.7meV, -0.5meV and 35.4meV. For clarity, a 34-nm-square field of view (FOV) is cropped from a larger 78nm*78nm FOV which we used to obtain the Fourier transfer images. **d-l**, Drift-corrected and symmetrized Fourier transforms of dI/dV conductance maps. White dots in **d** indicate the Bragg peaks. The spectral weight near center has been reduced by removing low frequency signal originating from defects. Dominant scattering vectors are indicated by q$_i$ (i=0, 1, 2, 3, 4). Dashed arrows and arcs denote their umklapp processes.

# Figure 3

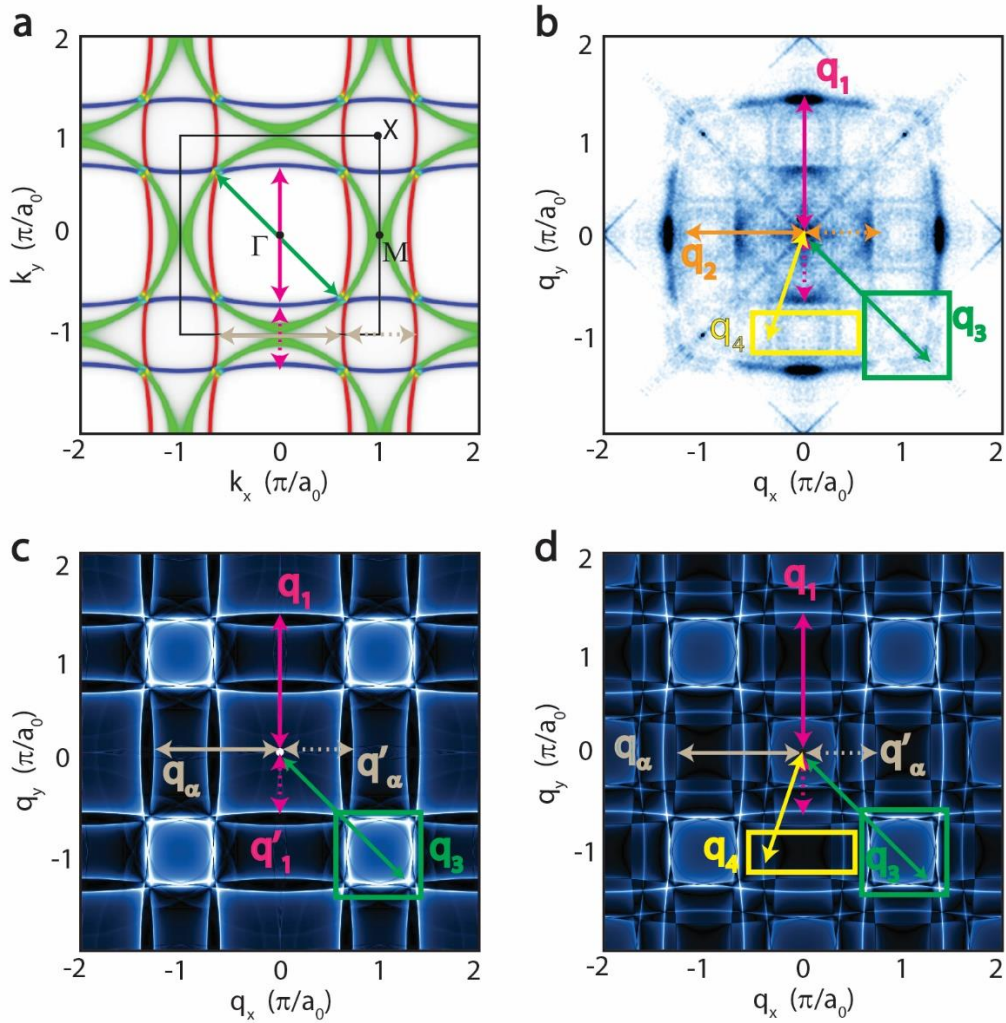

**Figure 3. Comparison of the FT-STS images with predicted QPI patterns**. **a**, Spectral density at Fermi level for the α, β and γ bulk bands. The relative contributions of each orbital to the bands are color-coded. Red, $4d_{xz}$; Blue, $4d_{yz}$; Green, $4d_{xy}$. Arrows with different colors denote the possible scattering. For a better view, we show CECs in the extended zone. **b**, QPI map at the Fermi energy (-0.5meV, same as in Fig.2h) and inequivalent Q- vectors are shown by arrows and colored squares. **c, d**, Theoretically simulated QPI patterns using T-matrix approaches for the original un-folded FSs (**c**), and including the folded replicas (**d**). $q_1$ and $q'_1$ represent intra-band scattering of bulk β band and its umklapp process, respectively, while $q_2$ and $q'_2$ for those of surface β band; $q_3$ represents a small portion of intra-γ-band scattering; $q_4$ represents scattering processes between unfolded bands and its folded replicas; $q_α$ and $q'_α$ in **c** and **d** represent intra-α-band scattering and its umklapp process.

**Figure 4**

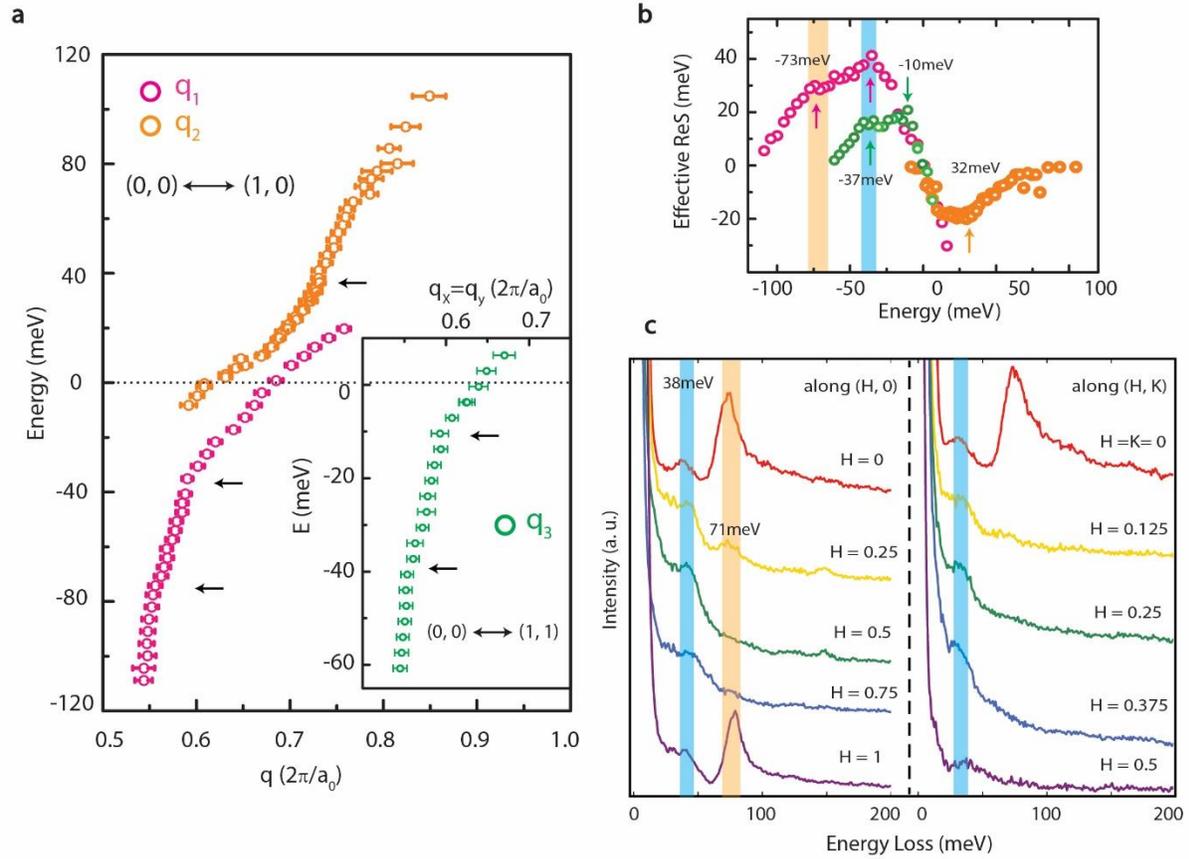

**Figure 4. Visualizing the electron-collective mode coupling in the Quasi-1D bands**. **a**, Dispersions of β band ($q_1$) and surface β band ($q_2$) extracted by fitting peaks in line-cut along ΓM direction. These peaks reflect the dynamic nesting processes involving band structures as well as the quasiparticle self-energy. Kink features are seen at energies of about -35meV and -70meV for $q_1$, and +32meV for $q_2$, as shown by arrows. Inset, Dispersion of $q_3$ along ΓX direction. An additional kink at -10meV is found. **b**, Corresponding effective real-part of quasiparticle self-energy (or ΔE) for the measured dispersion. A straight line connecting two points at $E_F$ and ±110meV (-60meV for $q_3$) in the dispersion is used as the 'bare' band for each **q** dispersion. Peaks at multiple energy scales are marked with arrows: -37 and -73meV for β band, 32meV for secondary β band and -11meV and -37meV for $q_3$. **c**, Momentum-resolved electron energy loss spectra taken at T = 100K. Two peaks at 38meV and 71meV are clearly revealed, which match the energy scales of kinks seen in QPI.

17